\documentclass[preprint]{aastex}
\usepackage{emulateapj5, psfig, epsfig}
\slugcomment{Accepted by The Astronomical Journal}

\shorttitle{Globular Clusters in the Sgr Stream}
\shortauthors{Bellazzini, Ferraro \& Ibata}

\def\ltsima{$\; \buildrel < \over \sim \;$}
\def\simlt{\lower.5ex\hbox{\ltsima}}
\def\gtsima{$\; \buildrel > \over \sim \;$}
\def\simgt{\lower.5ex\hbox{\gtsima}}
\def\kms{{\rm\,km\,s^{-1}}}

\def\kpc{{\rm\,kpc}}

\def\deg{^\circ}

\def\s{\ifmmode \widetilde \else \~\fi}
\def\={\overline}

\def\spose#1{\hbox to 0pt{#1\hss}}

\def\lta{\mathrel{\spose{\lower 3pt\hbox{$\mathchar"218$}}
     \raise 2.0pt\hbox{$\mathchar"13C$}}}
\def\gta{\mathrel{\spose{\lower 3pt\hbox{$\mathchar"218$}}
     \raise 2.0pt\hbox{$\mathchar"13E$}}}
\def\Dt{\spose{\raise 1.5ex\hbox{\hskip3pt$\mathchar"201$}}}    
\def\dt{\spose{\raise 1.0ex\hbox{\hskip2pt$\mathchar"201$}}}    

\def\dotsfill{\leaders\hbox to 1em{\hss.\hss}\hfill}

\def\Gyr{{\rm\,Gyr}}

\begin{document}

\title{Building up the globular cluster system of the Milky Way. \\
       The contribution of the Sagittarius galaxy}

\author{Michele Bellazzini, Francesco R. Ferraro}
\affil{INAF - Osservatorio Astronomico di Bologna, Via Ranzani 1, 40127, 
Bologna, ITALY}
\email{bellazzini@bo.astro.it, ferraro@apache.bo.astro.it}

\and

\author{Rodrigo Ibata}
\affil{Observatoire de Strasbourg, 67000 Strasbourg, France}
\email{ibata@newb6.u-strasbg.fr}


\begin{abstract}
We demonstrate  that there is a clear  statistical correlation between
the (X,Y,Z,$V_r$) phase-space distribution  of the outer halo Galactic
globular  clusters  (having $10\kpc  \le  R_{GC}\le  40\kpc$) and  the
orbital path of the Sagittarius dwarf spheroidal galaxy (Sgr dSph), as
derived by  Ibata \& Lewis.  At least  4 of the sample  of 35 globular
clusters  in this  distance range  were  formerly members  of the  Sgr
galaxy (at the  95\% confidence level), and are  now distributed along
the Sgr Stream, a giant  tidal structure that surrounds the Milky Way.
This is the first  instance that a statistically significant structure
associated with the Sgr dSph
has been detected  in the globular cluster population  of the Galactic
halo.  Together  with the four  well-known globular clusters  that are
located near the center of this tidally-disrupting dwarf galaxy, these
clusters  constitute $\simgt  20$\% of  the population  of  outer halo
($R_{GC} \ge 10\kpc$)  clusters.  The Sgr dSph was  therefore not only
an important  contributor to  the halo field  star population,  but it
also had a significant role in the building-up of the globular cluster
system of the Milky Way.
\end{abstract}


\keywords{(Galaxy:) globular clusters: general - Galaxy: halo - 
Galaxy: formation - Galaxy: structure - galaxies: dwarf}

{}{}
\section{Introduction}

The quest  to find ordered structures  in the system  of satellites of
the  Milky  Way  galaxy dates  back  to  30  years ago  when  possible
alignments  of  globular clusters  and/or  dwarf  galaxies along  wide
streams were first noted \citep{hm69,ly76,k77,k79,ly82}.  The mounting
consensus for  scenarios in  which the accretion  of satellites  has a
major  role  in  the  formation  of  the  outer  halo  of  the  Galaxy
\citep{sz78,z93} prompted a  new burst of such kind  of studies since the
mid  '90s  up to the present day \citep{maj94,lynlyn,f95,dana,palma}.   
Despite  the  many  interesting
suggestions,  none  of  the  quoted  studies was  able  to  provide  a
conclusive proof of  the reality of the alignments,  mainly because of
the overwhelming difficulty to  assess the statistical significance of
structures formed by inherently small numbers of objects.

However, recent theoretical and observational achievements may help us
to  look  into  the  problem   from  a  different  and  more  fruitful
perspective:

\begin{enumerate}

\item{} N-body simulations of the process of Galaxy assembly, starting from
        standard cosmological conditions (Cold Dark Matter - CDM), 
	strongly suggest that the hierarchical
	merging of satellites is the main driver of galaxy formation 
	\cite[see][and references therein]{DMsub,klyp}.

\item{} Independent of the assumed cosmology, it has been demonstrated that
        the accretion/disruption of satellites into the halo of a 
	larger galaxy may leave 
	long-lived relics in the form of streams of stellar (and/or dark
	matter) remnants that remain aligned to the orbital path of the
	parent satellite \citep{katy1,katy2,carb,orb,lucio}. 

\item{} Convincing observational evidence of the clumpy and ``filamentary''
        nature of the Galactic halo have been provided by many different groups
	\citep[e.g][]{sdss,vivas,dohm,yanny,ivez,carb,ami1,ami2,maj99}.

\end{enumerate}

The  {\em in  vivo}  example of  a  satellite accretion/disruption  is
provided  by   the  Sagittarius  dwarf   Spheroidal  galaxy  \cite[Sgr
dSph;][]{s1,s2}, which is currently merging with the Milky Way, and is
carrying its own globular cluster system (i.e., M~54, Ter~8, Arp~2 and
Ter~7, previously believed to  be normal Galactic globulars). There is
now clear  observational evidence that  the Sgr dSph is  loosing stars
under the strain  of the Milky Way tidal  field. These tidally-removed
stars are found along a huge (and quite coherent) stream extending all
over    the    sky     \citep[Sgr    Stream,    see][and    references
therein]{carb,dohm,david,sdss},  tracing  the   orbit  of  the  parent
galaxy.

\citet[][hereafter  IL98]{orb}, and  \citet{carb}  have simulated  the
evolution of  the Sgr  dSph over several  orbital periods (P  $\sim 1$
Gyr), computing  the orbit  of the galaxy  as well as  the phase-space
distribution  of  the debris  under  different  assumptions about  the
flattening of the CDM halo.  The initial conditions of the simulations
were based on  the known position and radial velocity  of Sgr dSph and
on its proper motion as estimated by \citet{s2,carb}.  The orbit has a
planar  rosette structure,  with  the  pole of  the  orbit located  at
[$\ell=90^\circ$,  $b=-13^\circ$] (i.e. a nearly  polar orbit), and  peri- and
apo-Galactic  distances of  $15\kpc$ and  $60\kpc$  respectively.  The
derived orbit has been successfully compared with the observed position
of  the  Sgr Stream  \citep{carb,orb-sds},  providing also  remarkable
indications that the dark halo of the Milky Way is nearly spherical.

In this  framework it is a  tantalizing application to  look for other
halo globulars that may be correlated with the orbital path of the Sgr
dwarf, and which  could be lying in the Sgr  Stream. In particular, we
look for the phase-space coincidence  of outer halo globulars with the
computed orbit of  the Sgr dSph from 1~Gyr ago up  to the present day,
searching  for the  most  recent episodes  of  globular cluster  loss,
i.e. the ones whose traces are most likely to be still detectable.

\section{Looking for structures}

For our  comparison we selected  from the catalogue  by \citet{harris}
the  35 globular  clusters  in the  range  of galactocentric  distance
$10\kpc  \le R_{GC}\le  40\kpc$. Among  these, 33  have  also measured
radial velocity  $V_r$. For sake of  brevity and clarity  we will call
this sample the Outer Halo  Sample (OHS), in the following.  With this
selection we avoid  the central part of the Galactic  halo where it is
less likely that  ordered structures can survive for  a long time, and
we leave  out of the sample  the handful of clusters  lying outside of
$R_{GC}\ge 60\kpc$, a region that lies beyond the Sgr Stream according
to the  IL98 orbit\footnote{There is a 28 kpc wide gap in the 
radial distribution of Galactic globular clusters 
\cite[see, e.g.][and references therein]{z85}. 
There are only five
clusters beyond $R_{GC}=40$ kpc, namely Pal~14, Eridanus, Pal~3, NGC~2419,
Pal~4, and AM1. Their respective galactocentric distances are $R_{GC}=$
65, 83, 90, 98, 99, and 117 kpc, much beyond the apogalacticon of the
Sgr orbit. The adopted outer radial threshold \cite[$R\le 40 \kpc$, quite 
similar to the one adopted by][i.e., $R_{GC}\le 36 \kpc$]{palma} 
provides the selection of a homogeneous sample without significant gaps.}.  
The adopted  OHS {\em does not  include the known
Sgr globulars}, to avoid the  detection of the obvious signal of their
clustering around the center of the Sgr galaxy.

In Figure~1 we show the OHS clusters (small solid circles) and the Sgr
orbit  in   the  planes  formed  by   the  rectangular  Galactocentric
coordinates\footnote{This Galactocentric  coordinate system is defined
such  that  the origin  lies  at the  Galactic  Center;  at the  Solar
position, $(-8,0,0)$, the $Y$-axis points in the direction of Galactic
rotation;  while  the  $Z$-axis  towards  the  North  Galactic  Pole.}
($X,Y,Z$, in kpc)  and in the $R_{GC}$ [kpc]  vs.  $V_r$ [km/s] plane.
The large full circles are the known Sgr globulars, which we also show
in the  plots for completeness.   Note that these clusters  lie around
the end  of the orbit corresponding  to the present  time ($t=0$).  We
highlight (with  encircled solid circles)  six more clusters  that lie
remarkably close  to the  orbit in all  the considered  planes.  These
clusters are:  Pal~12 \citep[whose association  to the Sgr  Stream has
been already  established by][]{pal12,dav12}, 
NGC~4147,  NGC~5634, NGC~5053,
Pal~5 and  Ter~3.  Is this association  {\em real} or could  it be the
mere occurrence  of a chance  alignment?  Though chance  alignments in
the four-dimensional phase space  (X,Y,Z,$V_r$) are not expected to be
very likely,  the key  point is to  quantify the probability  that the
observed   structure  could   have  originated   from   a  statistical
fluctuation.  To do  this we will compare the  observed distribution -
and its phase space distance to the Sgr orbit - with synthetic samples
(having  the  same  dimension  as  the OHS)  extracted  from  a  model
representing an unstructured parent halo.

\begin{figure*}
\figurenum{1}
\centerline{\psfig{figure=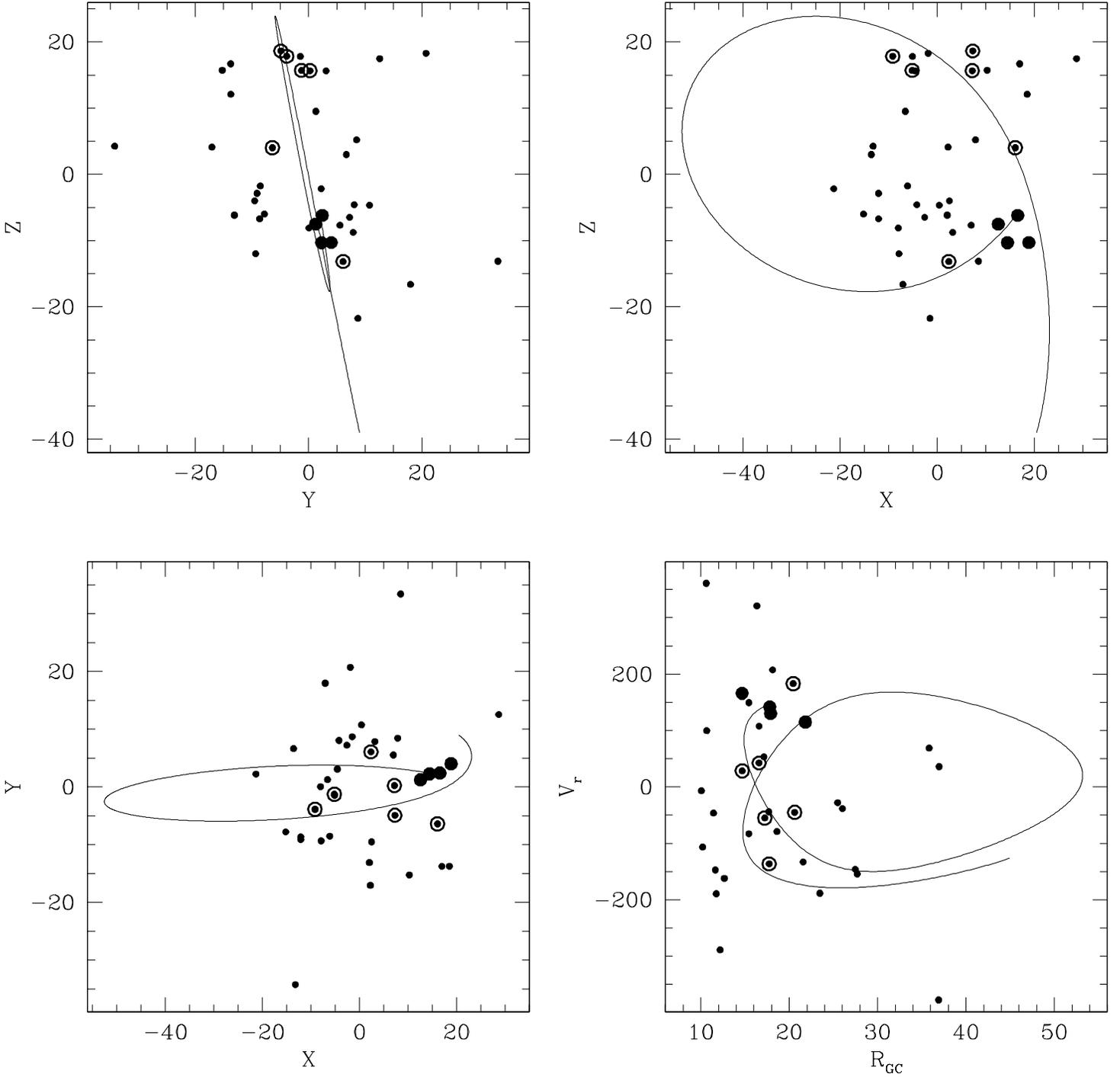}}
\caption{The distribution  of Galactic globular  clusters with $10\kpc
\le R_{GC}\le  40\kpc$ (OHS; small  full circles) in  the Y vs.   Z, X
vs. Z, X vs. Y, and $V_r$ vs. $R_{GC}$ planes.  The large full circles
are the  known Sgr  globulars M~54, Arp~2,  Ter~8 and Ter~7  while the
encircled points mark the clusters  that are near the Sgr orbital path
in all  of the planes  shown here (NGC~4147, Pal~12,  NGC~5634, Pal~5,
NGC~5053, and  Ter~3).  The  continuous line is  the orbit of  the Sgr
dSph since the present time (at the position of the Sgr clusters) to 1
Gyr ago, according to \citet{orb}.}
\end{figure*}

The most conservative comparison that can be made is with a model that
closely resembles the observed radial and velocity distribution of the
OHS.   Figure~2  (upper  panel)   shows  that  the  cumulative  radial
distribution of the  OHS is well reproduced by  a spherical halo model
with a density  distribution $\propto R^{-1.6}$.  A Kolmogorov-Smirnov
(KS) test  shows that the probability  that the OHS is  drawn from the
$\Phi \propto  R^{-1.6}$ model is  $\simeq 90$\%.  On the  other hand,
the  probability that  the same  sample is  drawn from  the  other two
models  shown  for  comparison  ($\Phi \propto  R^{-1.0}$,  and  $\Phi
\propto  R^{-2.5}$)  is  $\le  15$\%.   Doubts  may  be  cast  on  the
appropriateness of a spherical model.  It may be conceived that if the
{\em true} parent halo is  flattened, some excess of clustering of the
observed points  along an orbit with low  inclination may artificially
emerge in the comparison with  a spherical model.  This is clearly not
the case,  however, since the IL98  orbit is nearly polar,  i.e. it is
almost  perpendicular to the  Galactic Plane  (see Figure~1).   In the
lower panel of Figure~2 it  is shown that the observed distribution of
radial  velocity  of  the  OHS   is  well  reproduced  by  a  Gaussian
distribution    with    $<V_r>   =    -38\kms$    and   $\sigma_V    =
175\kms$. According  to a  KS test the  probability that  the observed
sample is drawn from the model distribution is $\simeq 90$\%.

In the following simulations we extract all the synthetic samples from
a   spherical   and  isotropic   model   with  $\Phi(R_{GC})   \propto
R_{GC}^{-1.6}$ and with the Gaussian distribution of radial velocities
shown in Figure~2.  For each simulated cluster (as well as for all the
OHS ones) we  computed the spatial distance from  the nearest point in
the Sgr  orbit ($D_{orb}$,  in kpc) and  the difference  between their
radial velocity and the one  predicted from the computed orbit at that
point ($\Delta V_r = V_r(obs) - V_r(orb)$).

\begin{figure*}
\figurenum{2}
\centerline{\psfig{figure=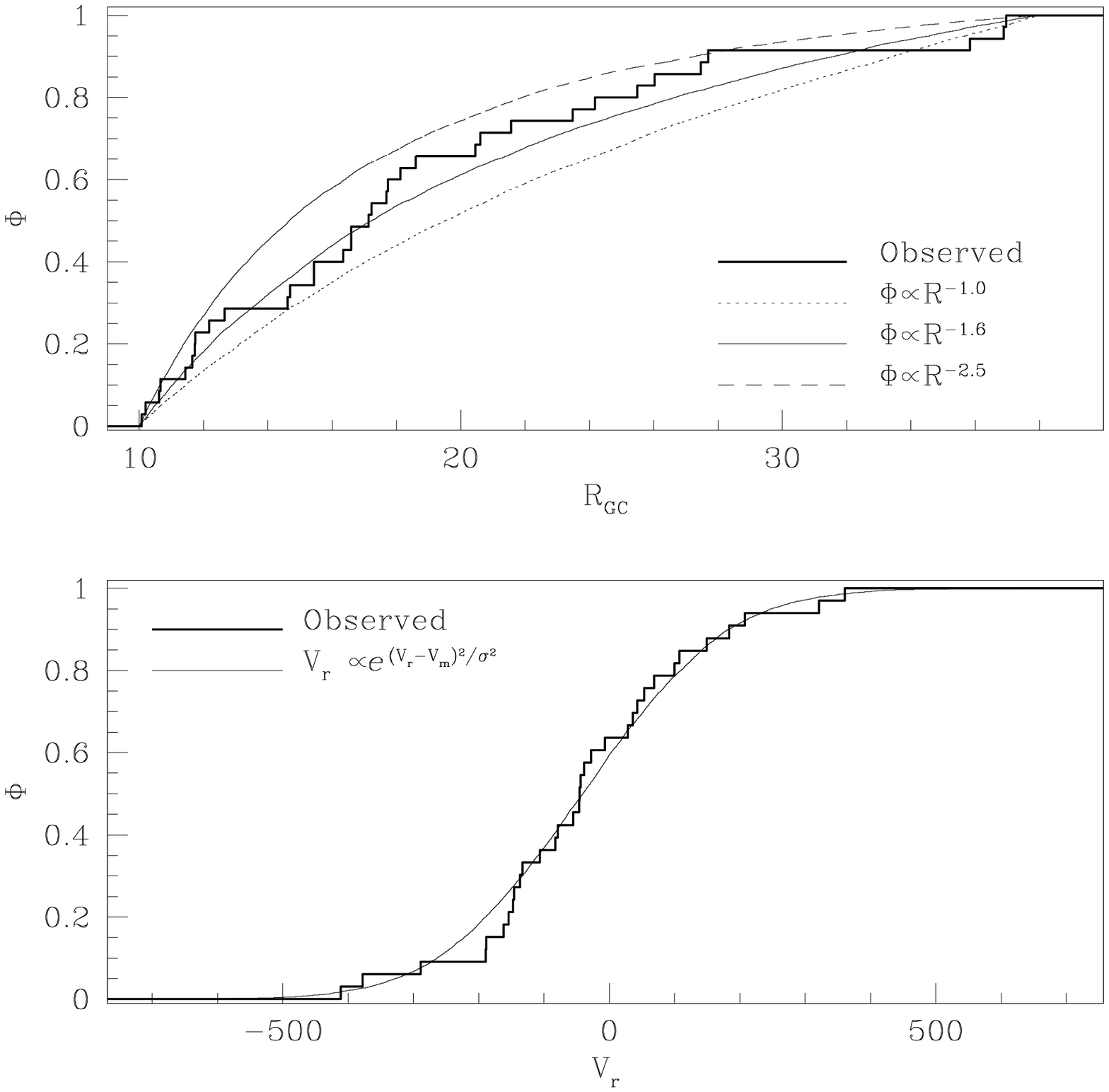}}
\caption{Upper panel:  the cumulative  radial distribution of  the OHS
globulars (thick continuous line) is compared to three different model
distributions in  the considered  radial range ($10\kpc  \le R_{GC}\le
40\kpc$).  According to a Kolmogorv-Smirnov test, the probability that
the  observed distribution  is drawn  from  the model  $ \Phi  \propto
R^{-1.6} $ is $P \simeq 90$\%,  while the other two models shown score
$P \le  15$\%.  Lower panel:  the observed cumulative  distribution of
radial  velocities for the  same clusters  (thick continuous  line) is
compared  with  a  Gauss  distribution  with  $<V_r>  =  -38\kms$  and
$\sigma_V  = 175\kms$.  Also in  this  case the  probability that  the
observed distribution is drawn from  the considered model is $P \simeq
90$\%.  }
\end{figure*}

In  Figure~3, the  $D_{orb}$ values  of the  selected  clusters (large
filled circles) are plotted  against their $\Delta V_r$. The encircled
points are the six clusters highlighted in Figure~1. A sample of 10000
synthetic  clusters (dots) extracted  from the  adopted model  is also
shown,  for comparison,  in  the  upper panel  of  Figure~3.  The  OHS
clusters  show a remarkable  over-density toward  the Sgr  orbit, that
lies in  the origin of the  axis in the considered  plane.  The dashed
dotted  lines enclose  the points  whose observed  radial  velocity is
within $\pm 60\kms$ of the velocity predicted by the IL98 orbit.  Note
that the expected velocity dispersion  of the Sgr debris along the Sgr
Stream  is $\sigma  \sim 60\kms$,  according to  \citet{orb-sds}.  The
continuous vertical  segments are placed  at $D_{orb}= $6, 12,  and 18
kpc.

The lower panel  of Figure~3 is arranged in the same  way, but in this
case  the  dots  represent  the  distribution of  the  points  in  the
best-fitting \citet{carb} simulation that  retains a bound core to the
present day. The boldface dots  are the particles that remain bound to
Sgr {\em or  were bound less than 3 Gyr ago},  while the ordinary dots
are  particles that  were  already  unbound at  that  time.  Note  the
remarkable similarity  with the distribution of the  OHS clusters.  In
particular, the  correlation between the particles that  have flown in
the  Sgr Stream in  recent times  (less than  3 Gyr  ago) and  the OHS
clusters  with $|\Delta  V_r|  <  60\kms$ at  any  $D_{orb}$ is  quite
striking.

We define $N_{D<x}$ as the number of synthetic or real clusters having
$-60\kms < \Delta  V_r < +60\kms$ and $D_{orb}<x  \kpc$.  For example,
it can  be seen from  Figure~3 that there  are five OHS  clusters with
$|\Delta V_r| <  60\kms$ and $D_{orb} < 6\kpc$,  thus $N_{D<6}=5$.  In
the same way we find $N_{D<12}=10$ and $N_{D<18}=14$ from the observed
sample.   Now the  question is  {\em what  is the  probability  that a
sample  having  $N_{D<x}$  greater   or  equal  to  the  observed  one
($N_{D<x}^{obs}$) is  drawn from  the assumed unstructured  model?} To
answer  this  question  we  randomly  extracted 10000  samples  of  32
synthetic clusters  from the  assumed model, and  for each of  them we
measured $N_{D<6}$, $N_{D<12}$, and $N_{D<18}$.

\begin{figure*}
\figurenum{3}
\centerline{\psfig{figure=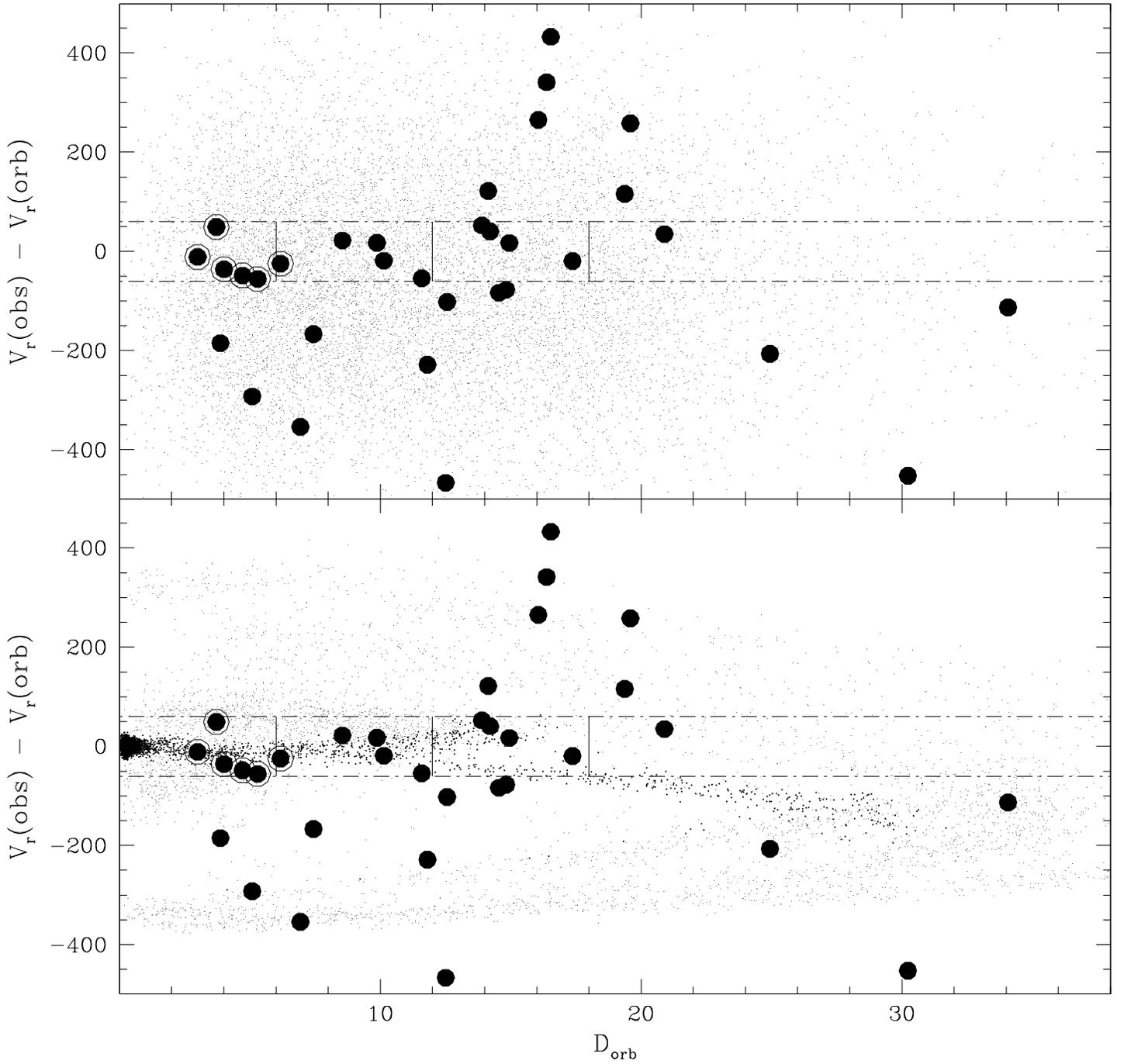}}
\caption{The  distribution  of the  OHS  globular  clusters about  the
predicted orbit of Sgr. The parameter $D_{orb}$ is the distance of the
cluster from the nearest point of the orbit, while $V_r(obs)-V_r(orb)$
is  the  difference  between  the  observed radial  velocity  and  the
prediction of the computed orbit  (at the nearest point in the orbit).
The observed  distribution (large full  circles) is compared  with the
distribution  of 10000  points randomly  drawn from  our  assumed halo
model (upper  panel) and  with the distribution  of the Sgr  debris as
computed  by  \citet{carb}  (lower  panel).   This  N-body  simulation
displayed in the  lower panel is the model  by \citet{orb} that gives
the  best fit  to the  positions and  velocities of  the  carbon stars
observed  in the  Sgr stream,  while retaining  a bound  center (their
dwarf  galaxy model  ``D1'', and  halo model  ``H2'', integrated  in a
Galactic potential with  circular velocity $v_c=200\kms$ at $50\kpc$).
The boldface dots show particles that remain bound  to Sgr or were
bound less than $3\Gyr$ ago, whereas the ordinary dots were already unbound
at that time.  Clearly, several  clusters in the OHS follow the
expected trend  of the recently  disrupted Sgr debris.   The encircled
points represent  the clusters similarly put in  evidence in Figure~1.
The  horizontal dashed-dotted  lines  enclose the  range $-60\kms  \le
\Delta  V_r \le 60\kms$,  while the  continuous vertical  segments are
located at $D_{orb} = 6$, 12 and 18~kpc.}
\end{figure*}

Figure~4  reports  the   distributions  of  $N_{D<6}$  (upper  panel),
$N_{D<12}$ (middle panel), and  $N_{D<18}$ (lower panel) for the 10000
simulated samples.  The respective  observed values are indicated by a
dashed  line.  In all  of  the considered  cases  the  mean and  modal
$N_{D<x}$  values  of  the  distribution  are much  smaller  than  the
observed ones. $N_{D<6}\ge 5$ occurs for only 200 simulated samples in
10000 (2\% of the cases), $N_{D<12}\ge 10$ occurs only for 42 samples
in 10000  ($0.4$\%), and $N_{D<18}\ge  14$ occurs only in  2 cases in
10000 ($0.02$\%).   It can be concluded that  it is highly improbable
that the observed clustering of the OHS globulars around the Sgr orbit
occur by  chance alignment. It is  important to remark  again that the
model from  which the  simulated samples have  been extracted  has the
same $R_{GC}$ and $V_r$ distribution of the observed sample. Thus {\em
the  Sgr  orbit appears  to  be a  strongly  preferred  subset of  the
phase-space  for the Galactic  globulars} in  the considered  range of
$R_{GC}$.

Finally, we  note that the  simulated samples that have  a significant
probability of  realization (e.g.,  $P\ge 20$\%) have  $N_{D<6}\le 3$,
$N_{D<12}\le  6$, and  $N_{D<18}\le 8$,  significantly lower  than the
observed values.  Hence, the  significance of the detected phase-space
structure cannot be due to the  actual correlation of just a couple of
clusters with the  Sgr orbit. According to the  distributions shown in
Fig.~4,  the  probability  that  $N_{D<18}\ge 10$  is  $P\simeq  5$\%.
Therefore it can be stated, with 95\% confidence, that at least 4 real
associations   are  needed   to   produce  the   observed  signal   of
$N_{D<18}=14$.

The  above test  demonstrates that  the observed  phase-space clumping
around the  orbit of the Sgr  dwarf galaxy is highly  unlikely to have
occurred  by  a  chance  coincidence,  if the  halo  globular  cluster
population  is indeed  distributed according  to the  simple spherical
model described above.  To address this concern, we performed a second
test, using ``bootstrapped'' artificial data sets constructed from the
real  OHS. The  artificial samples  were constructed  by  rotating the
position of  each real  globular cluster by  a random  azimuthal angle
about the Galactic center, and introducing a random flip perpendicular
to the  Galactic plane. This  implicitly assumes, of course,  that the
globular cluster distribution is  symmetric about the Galactic center,
as well as above and  below the Galactic plane.  A slight complication
arises from  the fact  that we do  not know  the proper motion  of many of the
clusters, so  we cannot deduce  the heliocentric radial  velocity that
the cluster  should have  at its new  position.  We therefore  have to
assume some model for the halo velocity distribution; we take a simple
Gaussian  model,  and draw  random  realizations  of  the total  space
velocity $\vec{V}$ consistent  with the observed heliocentric velocity
$V_r$ (that is, we  calculate the conditional probability of $\vec{V}$
given  $V_r$).   Having  defined  thereby the  3-dimensional  velocity
vector, we rotate the vector to the new position, and project it along
the line of  sight to the Sun (corrections for  the peculiar motion of
the Sun  and for Galactic rotation  are made).  With  a Gaussian model
that has $\sigma=175\kms$ and $V_r=-38\kms$, we find that out of 10000
artificial data sets, $F(N_{D<6} \ge  5) = 399$, $F(N_{d<12} \ge 10) =
102$, and $F(N_{D<18} \ge 14) =  18$. For comparison, for a model with
$\sigma=110\kms$ (or $\sigma=150\kms$) and $V_r=0\kms$, the statistics
are as follows: $F(N_{D<6} \ge 5)  = 463 (425)$, $F(N_{D<12} \ge 10) =
140 (121)$,  and $F(N_{D<18} \ge 14)  = 35 (28)$, where  the result in
brackets refers to the higher velocity dispersion model.  According to
these  tests using  ``bootstrapped''  random data  sets, the  observed
phase-space clumping is highly unlikely to have occurred by chance, in
agreement  with  our  previous  test.   The  fact  that  the  observed
structure  has  a slightly  lower  statistical  significance with  the
``bootstrapped'' samples than was  deduced from the previous test with
its  perfectly   isotropic  and  isothermal  random   samples  is  not
surprising.   This is  presumably a  consequence of  occasional chance
reappearances  of (part  of)  the structure  of  the real  OHS in  the
``bootstrapped'' samples.

\begin{figure*}
\figurenum{4}
\centerline{\psfig{figure=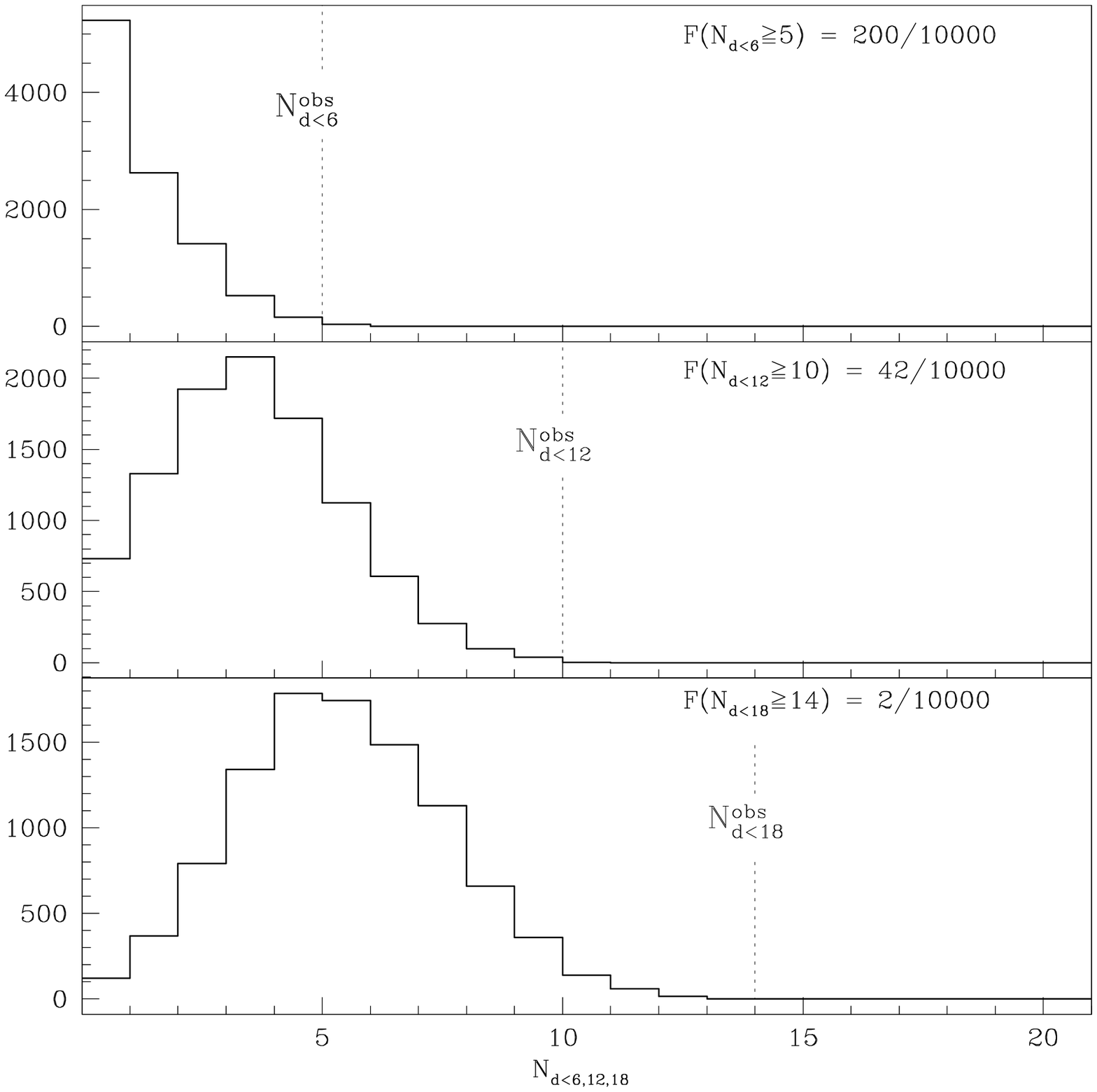}}
\caption{Results from the random drawing of 10000 synthetic GC samples 
(equivalent to the observed one) from the assumed halo model.}
\end{figure*}

As  a final  remark, we  note that  the region  of the  $\Delta  V_r -
D_{orb}$ plane  with $|\Delta V_r|  < 60\kms$ and $D_{orb}<18$  kpc is
also  populated by  particles  that  become unbound  more  than 3  Gyr
ago. Thus it has to be considered the possibility that some of the OHS
clusters  found in that  region drifted  into the  Sgr Stream  in more
ancient times.  For instance, this can be the case for NGC~5634, as we
argue elsewhere  \citep{n5634}. On the  other hand, the  discussion of
the  individual  OHS clusters  is  beyond  the  scope of  the  present
analysis.

\subsection{Proper Motions}

Thanks  to the  painstaking  effort  of a  few  teams of  astronomers,
estimates  of  the  proper  motions  (hereafter PMs)  of  41  Galactic
globular clusters are now  available \cite[see][for complete lists and
references]{dana,palma}.  Most  of these measures  have quite sizeable
uncertainties, ranging from  $\sim 0.2 $ mas/yr to  $\sim 2$ mas year.
Moreover,  subtle -  and unaccounted  - systematics  may  still affect
them,  as  suggested  by  the large  differences  between  independent
estimates  of  the  motion  of  the  same cluster.   In  the  list  by
\citet{palma} there are 14 cluster with more than one PM estimate. The
average absolute  difference among independent estimates  for the same
cluster are  $1.72\pm 1.29$  mas/year in $\mu_{\alpha}cos  \delta$ and
$2.90\pm 2.50$  mas/year in the  $\mu_{\delta}$ component. Differences
much larger than the quoted errors  (up to $\sim 10$ times) are common
and a  reasonable assumption of  the {\em minimum} uncertainty  of the
proper  motion  estimates  is  probably  $\sim  1$  mas/year  in  each
component.   Hence, while  available  PMs may  be  very valuable,  for
instance, to characterize the {\em  kind} of orbit followed by a given
globular, they are expected to have a modest constraining power in the
application presented  in this paper. Such constraining  power is also
lowered by the  fact that PMs predicted by the  model we are comparing
with  carry their  own, non  negligible, uncertainty  (typically $\sim
0.85$ mas/yr in  each component, which we estimate  from the intrinsic
dispersions in  distance, position and  velocity in the  N-body stream
models, assuming a 20\%  uncertainty in the Galactic circular velocity
at $50\kpc$).  Radial velocities, being unaffected by uncertainties in
the distance, are more reliably predicted by the model and are easily,
and much more accurately, measured.  For these reasons we preferred to
limit the bulk of our analysis to the the (X,Y,Z,$V_r$) phase-space.

\begin{figure*}
\figurenum{5}
\centerline{\psfig{figure=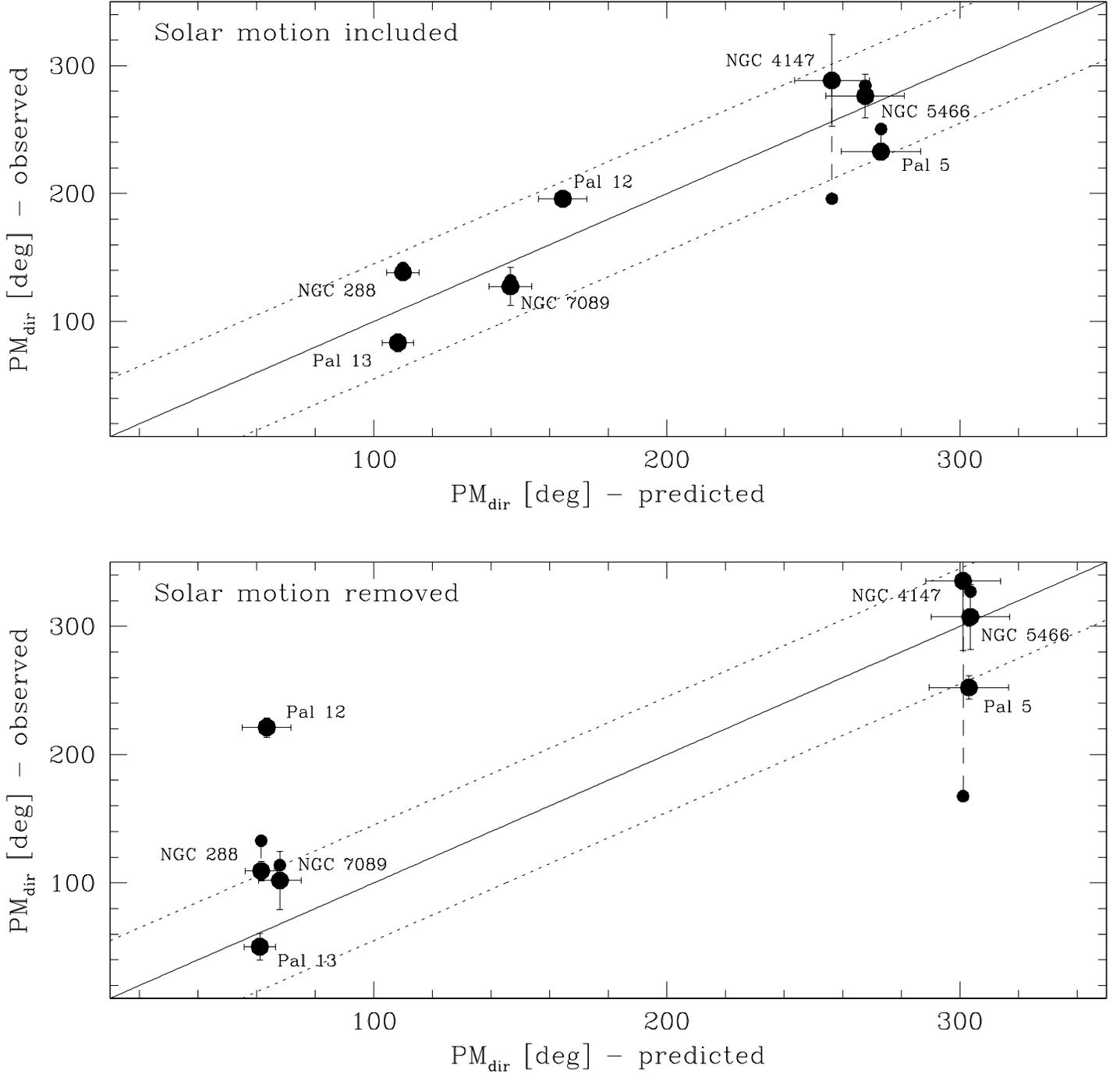}}
\caption{The predicted  direction of proper motions  are compared with
the observed ones  for the subset of candidate  Sgr Stream cluster for
which  estimates of proper  motion are  available \cite[from  the list
by][]{palma}.  When two  independent  estimates of  proper motion  are
available for  the same cluster we  report both: one as  a large filled
circle and the other as a  small filled circle, and the two symbols are
joined  by a  long dashed  line. The  large dot  of Pal~5  is  the 
estimate quoted by \citet{palma}  as Cudworth et al.   (2001), the small
dot is from \citet{shweitz}.  
The solid  line is the
locus of  perfect agreement  between predictions and  observation, the
dotted lines are displaced by $\pm 45$ deg. The error-bars are derived
from  the propagation  of the  uncertainties  in the  PM estimates  of
clusters as reported by \citet{dana,palma} and by the uncertainties in
the model  predictions, as described in \S2.1.   Panel (a): comparison
between  PMs including the  S.r.m. Panel  (b): comparison  between PMs
corrected for the S.r.m.}
\end{figure*}

Yet it  may be interesting  to compare the  PMs predicted by  the IL98
model for  the clusters that are  candidate members of  the Sgr Stream
(listed in Table 1, see below), with the observational estimates.  The
orientations (directions)  of the PM  vectors are expected to  be less
sensitive to systematics with respect to the actual moduli.  In Fig.~5
we  compare the  predicted and  observed  {\em directions}  of the  PM
vectors of the seven clusters  listed in Tab.~1 for which PM estimates
are available.  Four of these  seven clusters have two  independent PM
estimates: in  these cases we  report the comparison with  both values
(see caption). We stress that this kind of comparison {\em cannot have
a  serious impact  on  our  previous conlusions}  except  if both  the
following conditions were simultaneously fulfilled: (a) if accurate PM
estimates were available for all the considered clusters (or the large
majority of  them), and (b)  if less than  3-4 clusters were  found to
have observed PMs  in agreement with the predictions.  This case would
imply that  the correlation  among OHS clusters  and the Sgr  orbit we
detect  in the  (X,Y,Z,$V_r$) phase-space  is  largely due  to a  very
unlikely (but not impossible) chance alignment in four dimensions.  On
the other hand some (up to $\sim 10$, actually) of the clusters listed
in Table 1 {\em are expected not to belong to the detected structure}.
   
In panel (a) the predicted proper motions include a correction for the
Solar  reflex  motion (hereafter  S.r.m.),  assuming a  Galactocentric
distance of $8\kpc$, a circular velocity of $220\kms$ and the peculiar
motion  of the  Sun as  derived by  \citet{dehnen}.  The  predicted PM
directions of all the seven clusters agree with the observed values to
within $\pm  45\deg$. The  adoption of alternative  PM estimates  does not
change significantly the observed correlation.  However, the result of
this comparison may  be affected by the inclusion  of the Solar reflex
motion.  Hence,  in panel  (b) we present  the comparison  between the
plain (uncorrected  for the S.r.m) model predictions  and the observed
PM  directions after  correction  for  the S.r.m.  The result is very 
similar to that presented in panel (a) despite of the larger errors 
involved. The only cluster whose predicted and observed PM directions 
differ by more than $45 \deg$ is Pal~12, e.g. the only one for which 
independent evidences of membership to the Sgr Stream are already 
available \citep{dav12}.

In conclusion, it is found  that the available proper motion estimates
are fully compatible with the results presented in \S2.  Given the high
degree of correlation among predicted and observed PM directions shown in 
both panels of Fig.~5 one may also be tempted to draw firm conclusion
on the actual membership of individual clusters. Nevertheless, in our 
view, this may still be an hazard.  
If one looks at
the amplitude of  the individual corrections it is  found that (a) all
the considered clusters  have at least one component  of the PM vector
that is corrected by more than 50  \% of the observed value, (b) 4 out
of seven clusters  have at least one component  whose S.r.m correction
is larger that  100 \% of the observed values  and, (c) corrections as
large as  200 - 400  \% are indeed  applied in some cases.   Given the
remarkable  uncertainties in  the observed  PMs and  the uncertainties
involved in the S.r.m correction it is easy to conclude that currently
measured  PMs  are  not good  tracers  of  the  actual motion  of  the
considered  clusters and  are  instead predominantly  measures of  the
Solar reflex motion.  These considerations fully support the approach 
followed in \S2, e.g. using only positions and radial velocity  measurements 
in the search for statistical structures in the globular cluster system of 
the Milky Way.

\subsection{Comparison with previous analysis}

The  possible association  of  Galactic globulars  with  the Sgr  dSph
galaxy has  been considered  before \citep{irwin, dana,  pal12, pal13,
palma}. All these studies  strongly rely on PM estimates.  Furthermore
they  are centered  on testing  the actual  association  of individual
clusters to  the Sgr galaxy and  its remnants.  On the  other hand our
approach  and  our  results  are  statistical in  nature.  We  find  a
statistically  significant clustering  of OHS  clusters along  the Sgr
orbit  but we  have  no possibility  to  check individual  membership,
except for  the limited test  shown in \S2.1.   We can only  provide a
list of  possible members, ranked  according to their distance  to the
Sgr orbit in the  (X,Y,Z,$V_r$) phase-space (e.g., Tab.~1). Hence, the
comparison with the above quoted studies can be made only in a broadly
general sense.

\citet{dana} performed a thorough study  of the orbits of the Galactic
globulars for  which PM estimates  are available. In doing  this, they
identify three metal-poor clusters  having orbits typical of the thick
disc  (instead that  of  the halo,  as  expected) and  they check  the
hypothesis  that  the  peculiar   orbits  of  these  clusters  (namely
NGC~6254,  NGC~6626  and  NGC~6752)  could  be  due  to  their  former
membership  to  the Sgr  dSph,  concluding  that  this possibility  is
unlikely. All  these clusters have  $R_{GC}<10$, and are thus  are not
included in the  OHS considered here and we  cannot comment further on
the \citet{dana} results.

Following the  suggestion of \citet{irwin},  \citet{pal12} modeled the
(PM based) orbits of the Sgr dSph and of Pal~12 and concluded that the
available observations  are compatible  with the possibility  that the
cluster  was torn  from the  galaxy  $\sim 1.4  - 1.7$  Gyr ago.   Our
analysis confirms this result.

\citet{pal13} obtained  an estimate of  the PM of Pal~13  and compared
the  general characteristics of  the integrated  orbit of  the cluster
(total  energy,  angular   momentum,  eccentricity,  perigalactic  and
apogalactic radii)  to those of  five satellite galaxies of  the Milky
Way, including Sgr. They conclude  that a common origin for Pal~13 and
Sgr is  unlikely. However we note  that (a) Sgr provides  (by far) the
best match  to the  orbital parameters of  Pal~13 with respect  to any
other  considered galaxy  (see Siegel  et al.'s  Tab.~8),  the maximum
discrepancy  being $\sim  40$\%  in angular  momentum,  and (b)  the
orbital  parameters are provided  without any  uncertainty, as  if the
adopted  PMs were  perfectely  known.  In  our  view this  shortcoming
seriously weakens the conclusion by  these authors.  Here we adopt the
same PM  estimate obtained  by \citet{pal13}, reaching  the conclusion
that  Pal~13  is  a  possible  member  of the  Sgr  Stream  by  direct
comparison with a model that includes the Sgr Stream itself.

\citet{palma}  (hereafter   PMJ02)  performed  a   global  search  for
structures made of Galactic  satellites, adopting the technique of the
``orbital poles'',  originally introduced by  \citet{lynlyn}.  In this
kind  of  analysis one  searches  for  intersections  among the  great
circles described by the familiy of possible poles of the orbit of any
given  satellite.  As  stated by  PMJ02  ``any set  of objects  having
common origin and mantaining a  common orbit, no matter how spread out
in  the  sky,  will  have  great circle  pole  families  (GCPFs)  that
intersects  at the  same pair  of  antipodal points  on the  celestial
sphere''.  Adding the information from PMs, PMJ02 limit the valid pole
family of  each satellite  to an  arc of its  great circle  of orbital
poles (arc segment pole family, ASPF). With this technique they select
a  set of candidate-former-members  of the  Sgr galaxy  (Ter~7, Ter~8,
Arp~2,  M~54, M~53,  NGC~5053,  Pal~5, M~5,  Pal~12  and NGC~6356)  by
requiring that  their GCPF  come within  5 deg from  the Sgr  ACPF and
$6\le R_{GC} \le  36$ Kpc.  The first four clusters  in their list are
the well known present members of  the galaxy. They rise doubts of the
clear  association  of  each  of  the other  candidate  members.   For
instance, they  exclude NGC~5053 and M~53 because  their metal content
is  lower than  any of  the known  Sgr globulars\footnote{This  is not
completely true  since M~53  has the same  metallicity as  Ter~8, i.e.
$[Fe/H]=-1.99$  \cite[see][]{mont,taft}. See  also below,  for further
discussion about non-kynematic criteria  of selection of candidate Sgr
members.}.   With  the same  arguments  adopted  by \citet{pal13}  for
Pal~13 they  argue that Pal~5 and  Pal~12 are unlikely  members of the
family,  but  they cannot  rule  out  the  possibility. Finally,  they
conclude  that despite  the counter-arguments  presented  they mantain
their  original list of  candidates, with  the exception  of NGC~5053.
Elsewhere in  the paper, they also  note that NGC~5466  has energy and
angular  momentum  within  1-$\sigma$  of  those of  the  Sgr  galaxy.
NGC~6356, M~5 and M~53 have $R_{GC}  < 10$ kpc, hence are not included
in our sample. The possible association of  NGC~5466 with the Sgr dSph is 
confirmed by our analysis. The possible association of NGC~5053, Pal~5
and Pal~12 is also confirmed by the present work. PMJ02 also suggested
a  possible association  of NGC~4147  with a  group of  clusters whose
orbits  may be related  to those  of the  Magellanic Clouds,  but they
state that the result is  very uncertain.  Furthermore we note that on
the same basis the  possible association of M~53 (elsewhere classified
as a  candidate for  association with Sgr)  with the  Small Magellanic
Cloud  and Ursa Minor  is also  suggested by  PMJ02. We  conclude that
there is no serious disagreement  between the present analysis and the
one by PMJ02.

The present contribution, however,  {\em provides the first proof that
the orbit of Sgr is  a preferential subset of phase-space for globular
clusters inhabiting  the outer  halo of the  Milky Way}, and  we place
this result on a sound statistical basis, a result not accomplished by
any of  the previous studies. The  success of the  present analysis is
due  to  the  direct  comparison  of the  positional  and  kinematical
properties of the OHS clusters with  a realistic model of the orbit of
the Sgr dSph  and its relics, that has  been previously tested against
independent  observations \citep{carb,orb-sds}.  

Finally, we  shall shortly comment on  ``non-kinematic'' criteria that
have been used to assess the association of a given cluster to the Sgr
galaxy.      Many    authors    \cite[see,     e.g.][and    references
therein]{dana,palma} have  discussed the likeliness  of the membership
of  their  candidates  on  the   basis  of  the  similarity  of  their
metallicity,  Horizontal  Branch  (HB)  morphology  and/or  structural
parameters with those  of the four known Sgr  clusters. In our opinion
this approach may not be very  useful.  It is quite clear that Sgr was
a much  larger and complex system  in the past and  its present status
(as well as its present GC  system) may be not fully representative of
its original  range of properties.   If, by chance, the  cluster Ter~7
had been lost by the  Sgr galaxy in the previous perigalactic passage,
this kind of criteria would  have {\em erroneously} classified it as a
bad candidate for Sgr membership,  since Ter~7 is much more metal rich
($[Fe/H]\simeq -0.5$)  than any of  the other clusters that  are still
present in the  main body of Sgr dSph  ($[Fe/H]<-1.5$).  Moreover, the
existence of  population and/or metallicity gradients  may have driven
the  preferential  loss   of  metal  poor  populations  \cite[see][and
references   therein]{alard,sdss,n5634},   making   the   actual   low
metallicity limit of the  Sgr stellar population quite uncertain.  The
judgement  of the  likelihood of  membership on  the basis  of  the HB
morphology is even less justified since it is well known that clusters
of  different  morphologies  do  actually co-exist  in  real  galaxies
\citep[see,   e.g.,][and  references   therein,  for   the   case  for
Fornax]{fornax}.

A much more  reliable discriminant may be provided,  in the future, by
the  detailed abundance patterns  of the  clusters. For  instance, the
behaviour of $\alpha$ elements as a function of $[Fe/H]$ is determined
by  the star formation  history of  the parent  galaxy \citep[see][and
references therein]{mcw} that  was (probably) not the same  in the Sgr
dSph and in  the Galactic Halo.  In this context  it is interesting to
note that  Pal~12, Ru~106  \citep{brown} and Pal~5  \citep{smith} have
been found  to be less $\alpha$-enhanced than  Galactic Halo globulars
of similar metallicity.

For the above  reasons, we rely only on  phase-space parameters in our
analisys,  recalling that  by searching  the  lost relics  of the  Sgr
system we are trying the  reconstruct its original properties, not the
present day ones.

\section{Conclusions}

We  have  demonstrated that  there  is  a  coherent structure  in  the
phase-space  distribution of the  outer halo  globular cluster  of the
Milky Way, strongly  correlated with the orbital path  of the Sgr dSph
galaxy.   This correlation  cannot  have originated  by  chance, as  a
random realization of an unstructured  halo, if such halo has the same
distribution  of galactocentric  distance and  radial velocity  as the
considered OHS clusters.
  
Several of the OHS globular clusters have spatial positions and radial
velocities compatible with the hypothesis that
they are following the same orbit as the Sgr dSph, i.e.  they probably
belong  to the  Sgr Stream.   Furthermore  the spread  in phase  space
around the Sgr orbit of such clusters is similar to that predicted for
the Sgr  Stream population, according to the  numerical simulations by
\citet{carb}.
 
We conclude, with 95\% confidence, that {\em at least 4 (out of 32) of
the OHS  clusters are  physically associated with  the Sgr  Stream and
were former  members of  the Sgr dSph}.   It should be  noted however,
that the  analysis presented  in this contribution  is best  suited to
identify  clusters   that  became  unbound   relatively  recently.  In
principle, more OHS globular clusters could be associated with the Sgr
stream.

In  Table~1  we report  the  list  of  the selected  globulars  having
$|\Delta V_r|<60$  Km/s and  $D_{orb}<18\kpc$, in order  of increasing
$D_{orb}$. The  proper motions  predicted by the  IL98 model  are also
reported  in Table~1.  Since the  adopted analysis  is  statistical in
nature it is clear that some of the clusters listed in Tab.~1 may be just
occasional interlopers of the Sgr orbit (up to 10 over 14 in the worst
case, according  to our estimate).  The comparison with  the available
proper motion  estimates shown in \S2.1 does not provide any strong 
indication in this regard.  While  the
actual membership to the Sgr  Stream of each individual cluster can be
firmly established only with  very accurate proper motion measures (as
will be achieved by dedicated space  missions like GAIA or SIM), it is
reasonable to assume that the  clusters with the smaller $D_{orb}$ are
the ones for which the membership is more likely (see \S2.1).  We make
the hypothesis that the six  clusters more closely associated with the
Sgr orbital path  (i.e. those put in evidence in  Figures~1 and 3) are
former members of  the Sgr dSph.  In this case  the number of globular
clusters originally in the Sgr galaxy is $N_{cl}=10$.  If we adopt for
the  Sgr  dSph the  same  {\em  globular  cluster specific  frequency}
\citep[$S_N=N_{cl}10^{0.4(M_V+15)}$][]{hvdb}   of   the   only   other
Galactic  dSph  that  has  a  globular cluster  system,  i.e.   Fornax
$S_N\simeq 26$, we obtain an  estimate of the total absolute magnitude
of Sgr before the occurrence  of any tidal stripping, $M_V\simeq -14$.
\cite{sdss} estimated that  in the Sgr Stream there  are as many stars
as in  the present undisrupted  body of the  galaxy, thus the  {\em ab
initio} total  luminosity of  the Sgr dSph  was roughly two  times the
present value.   Hence the total  absolute magnitude at that  time was
$M_V-\log(2)\simeq -14.1$ \citep[where $M_V=-13.4$, from][]{M98}.  The
excellent  agreement  between the  two  independent  estimates of  the
initial  $M_V$  fully  supports   the  plausibility  of  the  proposed
scenario.

According to the  results presented, it emerges that  the Sgr dSph was
not   only   an   important    contributor   of   halo   field   stars
\citep{ivez,yanny,sdss}  but it  also had  a significant  role  in the
building-up of the globular clusters system of the Milky Way.

\acknowledgments

M.B. and F.R.F. acknowledge the  financial support to this research by
the italian {Ministero  dell'Universit\'a e della Ricerca Scientifica}
(MURST) through the grant p.  2001028879, assigned to the project {\em
Origin and Evolution of Stellar Populations in the Galactic Spheroid}.
This research has made use of NASA's Astrophysics Data System Abstract
Service.


\clearpage


\begin{deluxetable}{lcccccc}
\tablecolumns{7}
\tablewidth{0pc}
\tablecaption{OHS clusters in the Sgr Stream}
\tablehead{
\colhead{Name}& \colhead{D$_{orb}$ \footnotesize{[kpc]}} 
&\colhead{$\Delta V_r$ \footnotesize{[$km/s$]}} 
& \colhead{$\mu_{\alpha}cos ~\delta$\footnotesize{[mas/year]}} &
\colhead{$\mu_{\delta}$\footnotesize{[mas/year]}} 
&\colhead{ $\mu_{l}cos ~b$\footnotesize{[mas/year]}} & 
\colhead{$\mu_b$\footnotesize{[mas/year]}}\\ }
\startdata
       Pal 12   &     3.00   &	   -11.2  &  0.67& -2.41&-2.324 & -0.934 \\
     NGC 4147   &     3.72   &	    49.1  & -2.99& -0.73&-2.284 & -2.061 \\
        Pal 5   &     4.01   &	   -36.0  & -4.68&  0.25&-3.468 &  3.146 \\
NGC 5634\tablenotemark{a}& 4.72 & -48.6 & -4.64  &  0.25&-3.579 &  2.968 \\
     NGC 5053   &     5.30   &	   -55.7  & -3.78& -0.28&-3.754 &  0.494 \\
     Terzan 3   &     6.18   &	   -24.8  & -4.09& -0.29&-2.858 &  2.941 \\
     NGC 5466   &     8.55   &	    21.9  & -3.97& -0.16&-3.818 &  1.104 \\
      NGC 288   &     9.88   &	    17.2  &  3.37& -1.22& 1.235 & -3.368 \\
     NGC 7089   &    10.14   &	   -18.9  &  1.47& -2.23&-2.037 & -1.727 \\
     NGC 6426   &    11.59   &	   -54.4  & -4.15& -0.25&-2.891 &  2.981 \\
     NGC 5824   &    13.90   &	    52.5  & -4.34& -0.09&-3.023 &  3.119  \\
       Pal 13   &    14.21   &	    40.4  &  3.47& -1.14& 1.695 & -3.237  \\
        Pal 2   &    14.95   &	    16.9  &  3.99&  0.20& 2.933 &  2.707  \\
      Rup 106   &    17.37   &	   -19.4  & -4.67&  0.24&-3.439 &  3.176  \\
      &&&&\\
      &&&&\\
     NGC 6715   &     0.45   &      -4.2  & \nodata & \nodata& \nodata & \nodata \\
     Terzan 7   &     1.89   &      34.0  &\nodata  & \nodata& \nodata & \nodata \\
     Terzan 8   &     2.04   &      -2.0  & \nodata & \nodata& \nodata & \nodata \\
        Arp 2   &     5.13   &	    -26.6 & \nodata & \nodata& \nodata & \nodata \\
\enddata
\tablecomments{Distance  from the  nearest point  in the  orbital path
followed  by  the  Sgr  dSph  during  the  last  Gyr  ($D_{orb}$)  and
difference  in  radial velocity  between  the  considered cluster  and
prediction of the computed orbit  at that point ($\Delta V_r$) for the
OHS clusters having  $|\Delta V_r| < 60 $ km/s  and $D_{orb}< 18$ kpc.
The clusters are ordered with  growing $D_{orb}$. The known members of
the  Sgr  system are  also  reported  at the  end  of  the table,  for
comparison.  The  first six clusters of  the list are the  ones put in
evidence in  Fig.~1 and  Fig.~3. The four known members of the Sgr 
globular cluster system are also reported for comparison in the last four rows
of the table. The proper  motions predicted  by the
IL98 model [under  the hypothesis that (a) the  clusters belong to the
Sgr stream, and (b) they have  been lost during the last orbit of Sgr,
see \S2] are  also reported, in both the equatorial and galactic 
reference system. The intrinsic dispersion  of the modelled
stream,  combined with  estimated 20\%  uncertainties in  the Galactic
mass model, translate  to uncertainties of $\sim 0.85$  mas/yr on each
PM component.}
\tablenotetext{a}{\citet{n5634}  argue that NGC  5634 was  likely lost
before  the last  peri-centric passage  of Sgr,  so the  proper motion
predictions listed here may be substantially in error.}
\end{deluxetable}

\end{document}